\documentclass[twoside,a5paper,11pt]{article}
\usepackage{iwoc-a5}
\title{\Large \bf Recent results from the UrQMD hybrid model for heavy ion collisions}
\author{Marcus Bleicher, Stephan Endres, Jan Steinheimer, Hendrik van Hees}
\date{}
\begin{document}
\maketitle

\begin{center}
\vspace*{-0.3cm}
{\it  Frankfurt Institute for Advanced Studies (FIAS), Ruth-Moufang-Strasse 1, 60438 Frankfurt am Main, Germany\\
Institut f\"ur Theoretische Physik, Johann-Wolfgang-Goethe-Universit\"at, Max-von-Laue-Strasse 1, 60438 Frankfurt am Main, Germany\\
}
\end{center}

\vspace{0.3cm}

\begin{center}
{\bf Abstract}\\
\medskip
\parbox[t]{10cm}{\footnotesize These proceedings present recent results
  from transport-hydrodynamics-hybrid models for heavy ion collisions at
  relativistic energies. The main focus is on the absorption of
  (anti-)protons in the hadronic afterburner stage of the reaction,
  di-lepton production at SPS and heavy quark dynamics.}
\end{center}

\section{Introduction} 

A major theme in todays high energy heavy ion physics is to explore the
phases of Quantum Chromodynamics (QCD) at very high densities and
temperatures.  To connect ab-initio information on the properties of
QCD-matter, e.g., the Equation of State (EoS) or transport coefficients
like the viscosities with experimentally observable quantities one has
to rely on transport approaches that describe the time evolution of the
hot and dense matter created until the system has ceased to interact.
Transport models and hydrodynamic approaches have a long tradition in
providing this link.  Unfortunately, the areas of application of both
approaches seems mutually exclusive: Boltzmann equation based transport
simulations are well suited for the less dense stages of the reaction or
for lower energies, while hydrodynamic simulations are only justified
during the most dense stages of the reaction's evolution or at very high
collision energies.

\section{Model description} 

For the present studies we use the UrQMD model v3.3 in hybrid
mode\cite{Bass:1998ca,Bleicher:1999xi,Petersen:2008dd}.  This model
couples the fluctuating initial state \cite{Bleicher:1998wd} generated
event-by-event by the hadron and string dynamics from UrQMD to an ideal
hydrodynamic evolution. For the evolution of the hydrodynamic part
different equations of state can be applied, including a hadron gas EoS
and a chiral EoS with a transition to a quark-gluon plasma.  At the end
of the hydrodynamic evolution, defined by a transition energy density,
the hydrodynamic cells are converted to particles with a Cooper-Frye
prescription \cite{Huovinen:2012is}, and the decoupling stage is handled
by the UrQMD hadronic cascade \cite{Becattini:2012sq}.  For similar
approaches by other groups we refer
to\cite{Magas:2002ge,Csernai:2005ht,Andrade:2005tx,Hirano:2005wx,Nonaka:2005aj,Werner:2010aa}
and references therein.
\begin{figure}[t] 
\begin{center}
\includegraphics[width=0.45\textwidth]{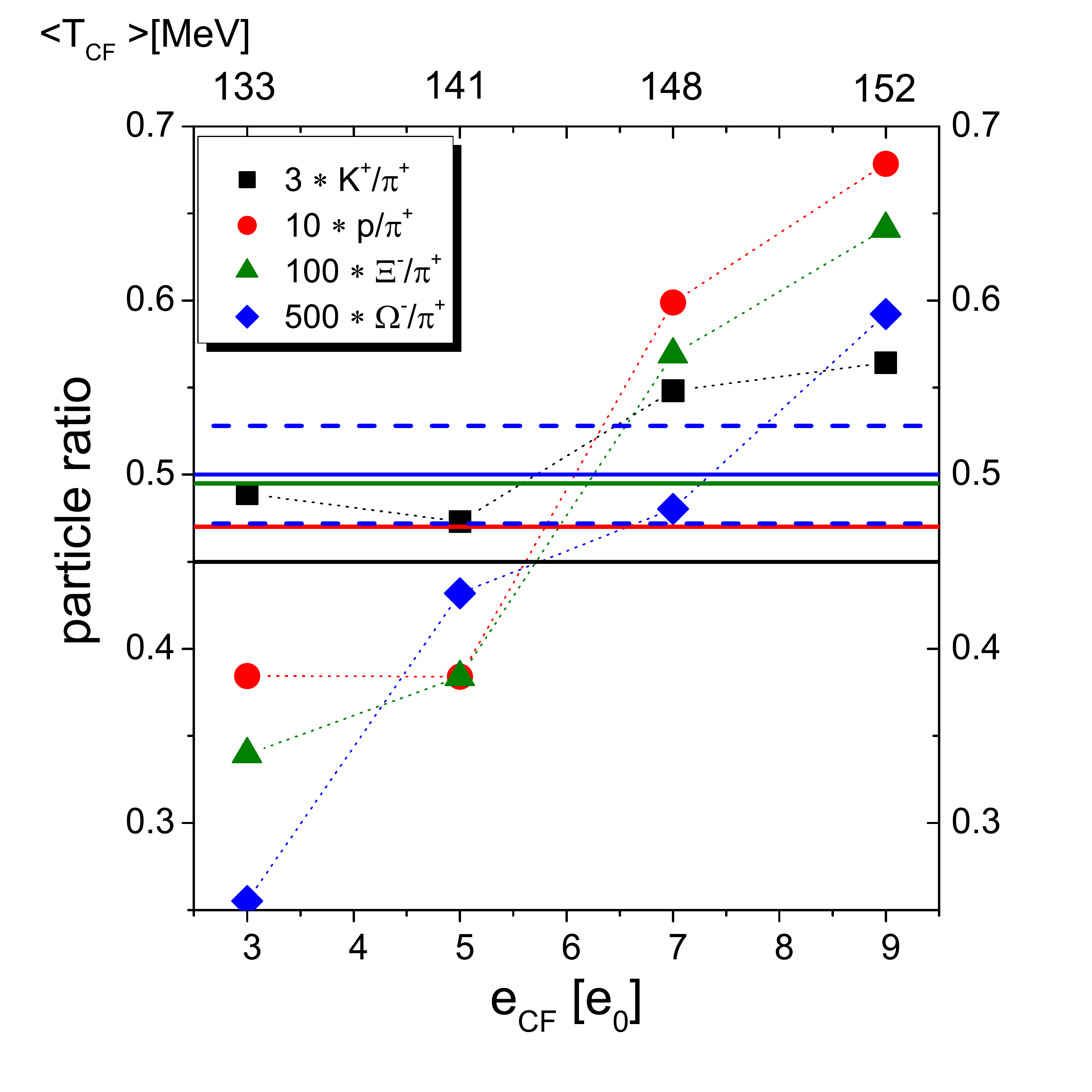}
\includegraphics[width=0.45\textwidth]{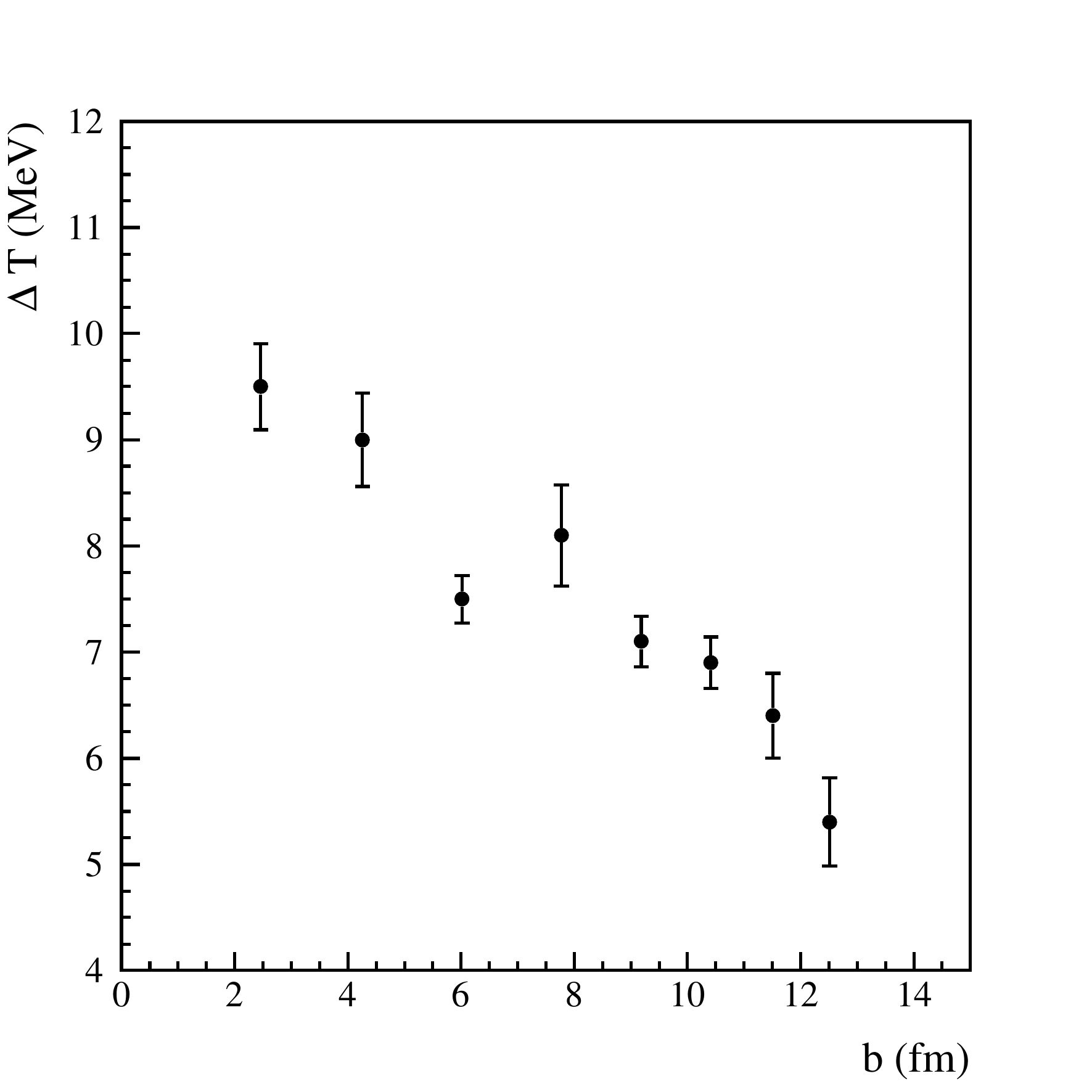}
\end{center} 
\vspace{-0.5cm} 
\caption{\label{fig1}Left: Modification of the final particle ratios as
  a function of the transistion energy density $\epsilon_{\mathrm{CF}}$ for the
  Cooper-Frye prescription in Pb+Pb collisions at
  $\sqrt{s_{NN}}=2.76$~TeV\cite{Steinheimer:2012rd}.  Right: Corrected
  temperature of the freeze-out points on the phase diagram
  \cite{Becattini:2012xb,Becattini:2014hla}.}
\end{figure}

\section{Matter and Antimatter} 

The effect of the hadronic afterburner can be seen directly in the
yields of the protons and anti-protons in nuclear collisions at the LHC.
Fig.\ref{fig1} (left) shows the modifications of the particle yields due
to the hadronic corona as a function of the energy density at which the
Cooper-Frye particlization is applied.  One observes that for realistic
transition energy densities around $\epsilon_{\mathrm{CF}}=5\epsilon_0$
the proton and anti-proton yields are reduced by approximately 50\% in
line with the observed values at the LHC \cite{Steinheimer:2012rd}.  A
similar conclusion is also reached if the back reaction
$5\pi\rightarrow p\bar p$ is included \cite{Pan:2014caa}.  The
systematic error, if the back reaction is neglected is on the order of
10\% \cite{Pan:2014caa}.  Fig.  \ref{fig1} (right) shows the
temperatures extracted from chemical fits for different centralities in
Pb+Pb reactions at $\sqrt{s_{NN}}=2.76$~TeV.  The temperature
differences $\Delta T$ are obtained from the temperature differences
between uncorrected fits and `corrected' fits (i.e., corrected for
baryon absorption in the hadronic corona as given by UrQMD) to the model
data \cite{Becattini:2012xb,Becattini:2014hla}.
\begin{figure}[t] \begin{center}
\includegraphics[width=0.45\textwidth]{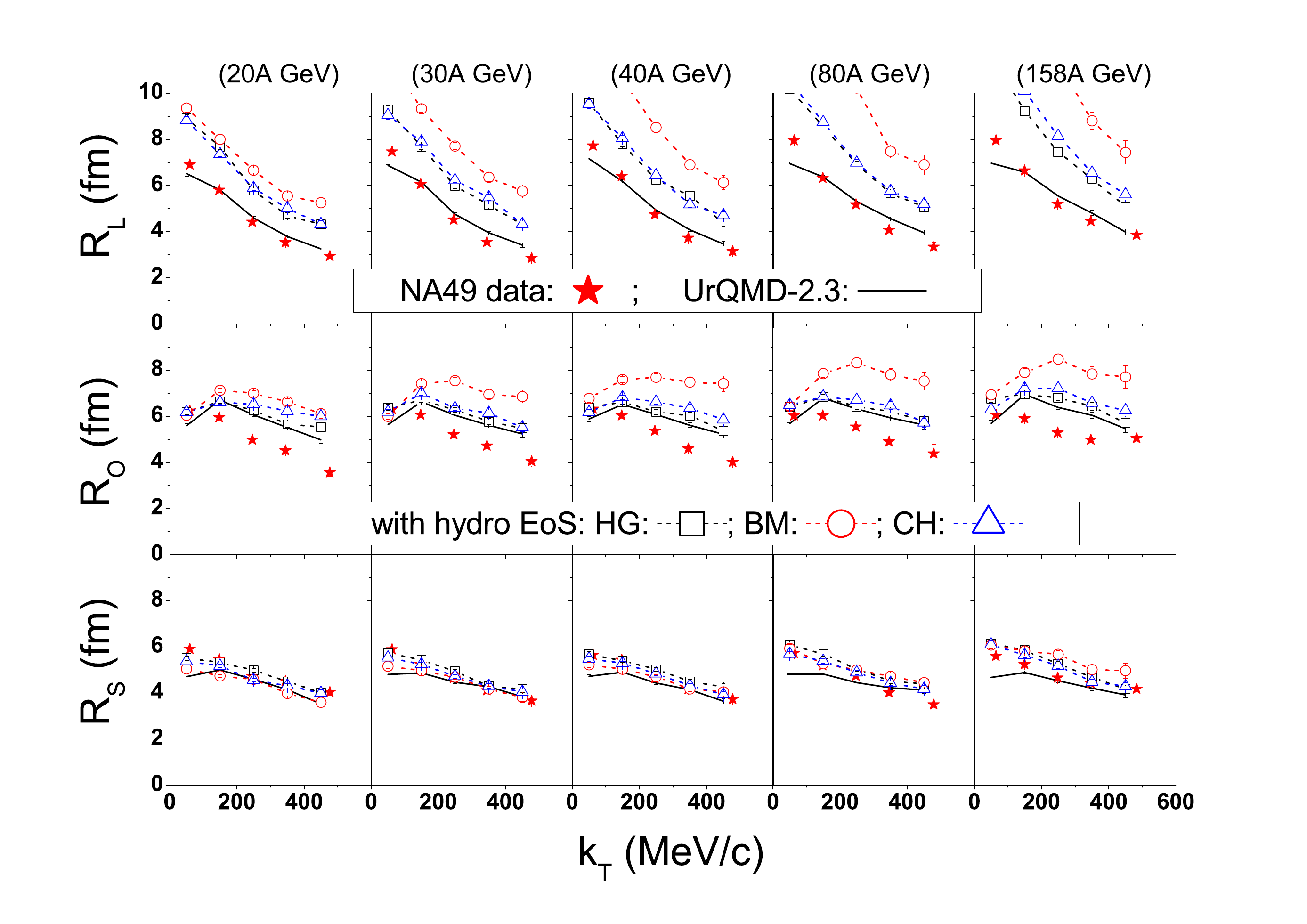}\includegraphics[width=0.45\textwidth]{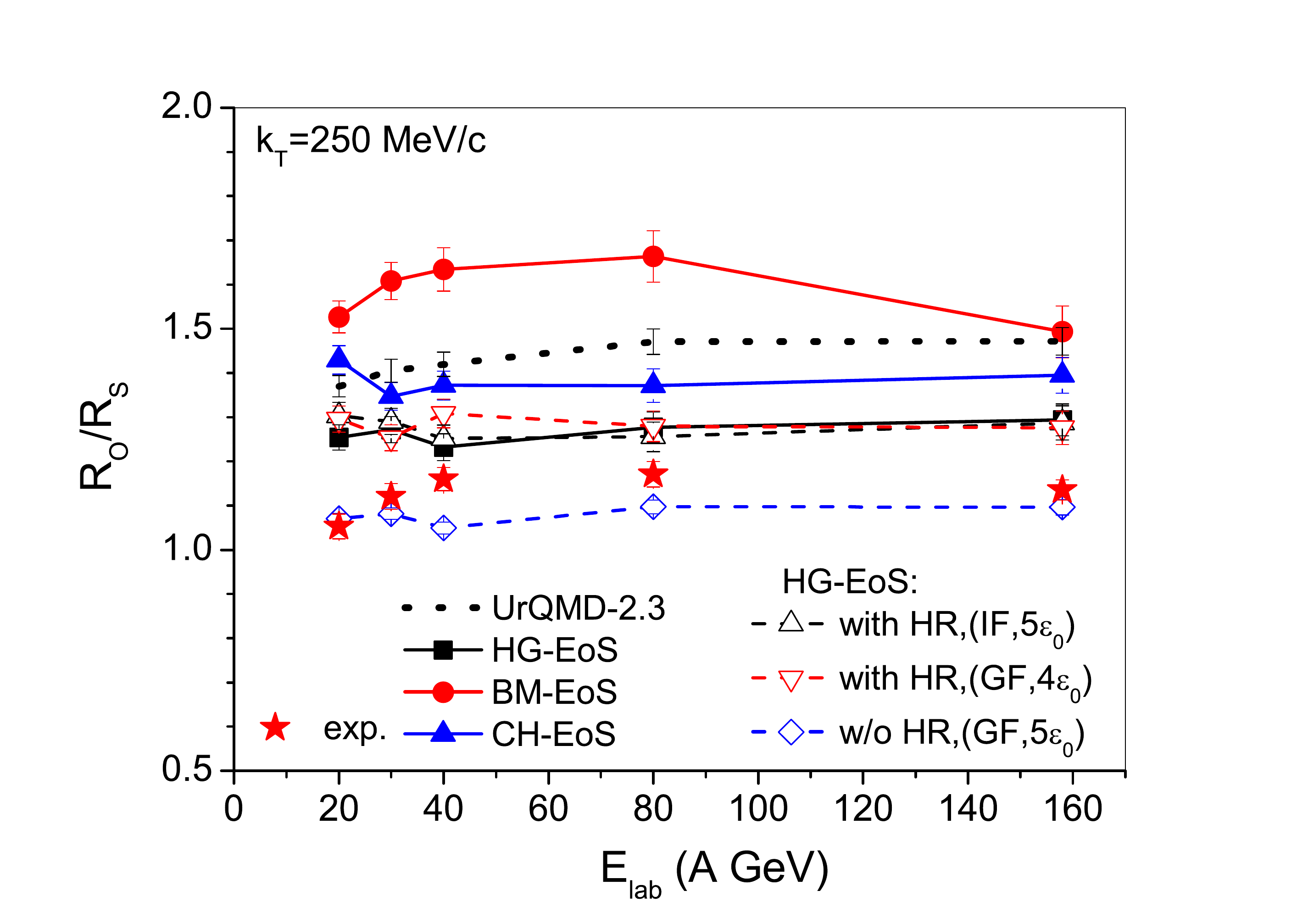}
\end{center} \vspace{-0.5cm} \caption{\label{fig2}Left: HBT radii for
different beam energies as a function of the transverse momentum of the pion
pair for different equations of state \cite{Li:2008qm}.  Right: Ratios of
the out to side radii as function of beam energy and for different equations
of state \cite{Li:2008qm}.}
\end{figure}

\section{Hanburry-Brown--Twiss Correlations} Let us next explore the
effect of different equations of state.  One expects that a phase
transition to a quark-gluon plasma (and back to the hadron gas) should
result in a delay of the expansion of the system depending on the
magnitude of the latent heat.  To explore this effect we compare hybrid
simulations for different equations of state using Hanburry-Brown--Twiss
(HBT) correlations \cite{Li:2008qm}.  Fig.  \ref{fig2} (left) provides a
comparison of various equations of state (hadron gas EoS, chiral EoS,
bag model EoS, and a pure UrQMD cascade simulation).  One observes that
the pure UrQMD simulation, the hadron gas EoS and the chiral EoS provide
a reasonable description of the data, while the bag model EoS clearly
overshoots the data.  The data and the simulations are summarized in
Fig.  \ref{fig2} (right) for the $R_{\mathrm{out}}/R_{\mathrm{side}}$
ratio for different beam energies.  Here one clearly observes the
expected maximum in the life time of the system in case of the bag model
EoS.  However, the data seem to favor a transition with a small latent
heat without a substantial time delay as provided by the hadron gas EoS
and the chiral EoS.
\begin{figure}[t] \begin{center}
\includegraphics[width=0.45\textwidth]{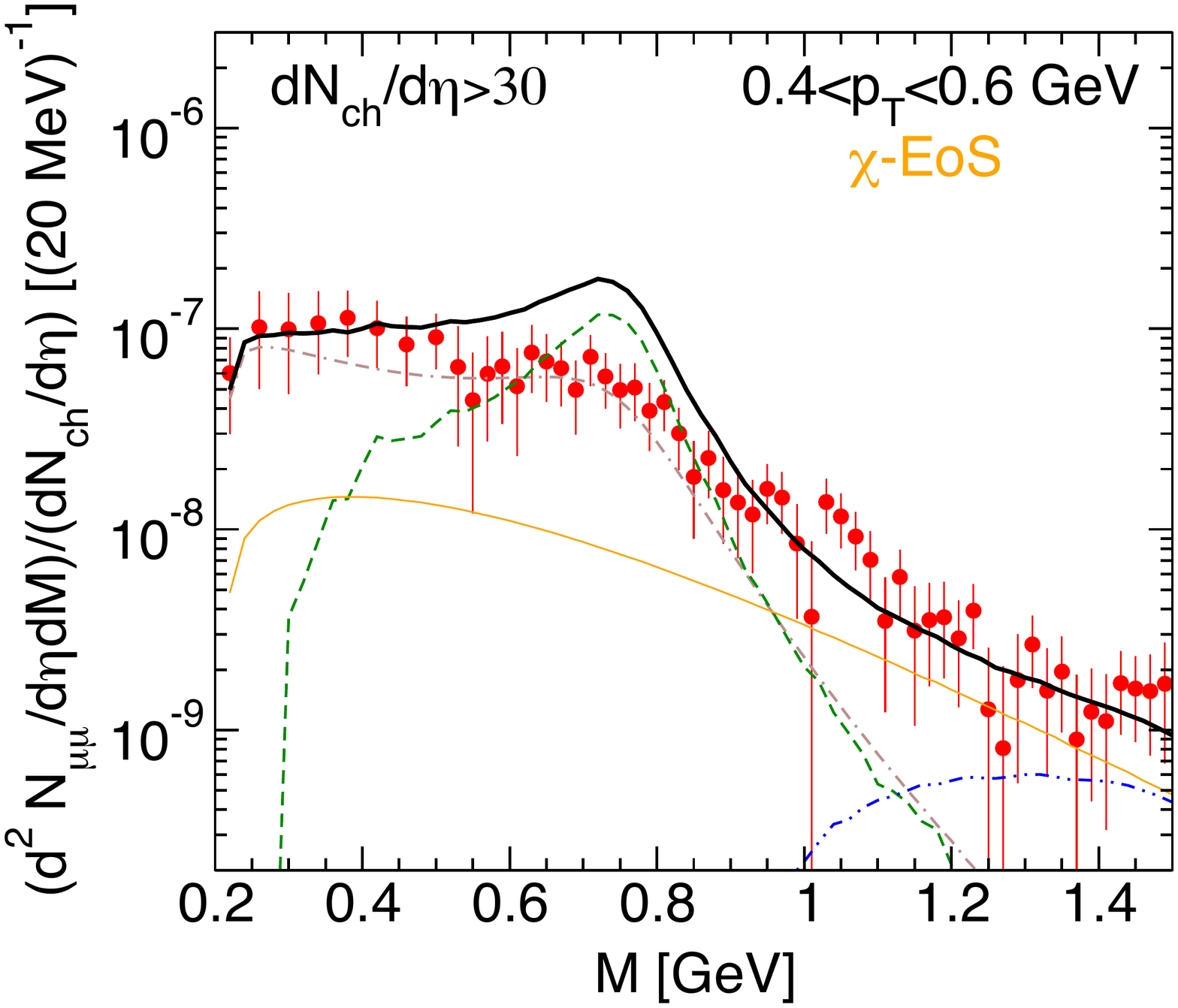}\includegraphics[width=0.45\textwidth]{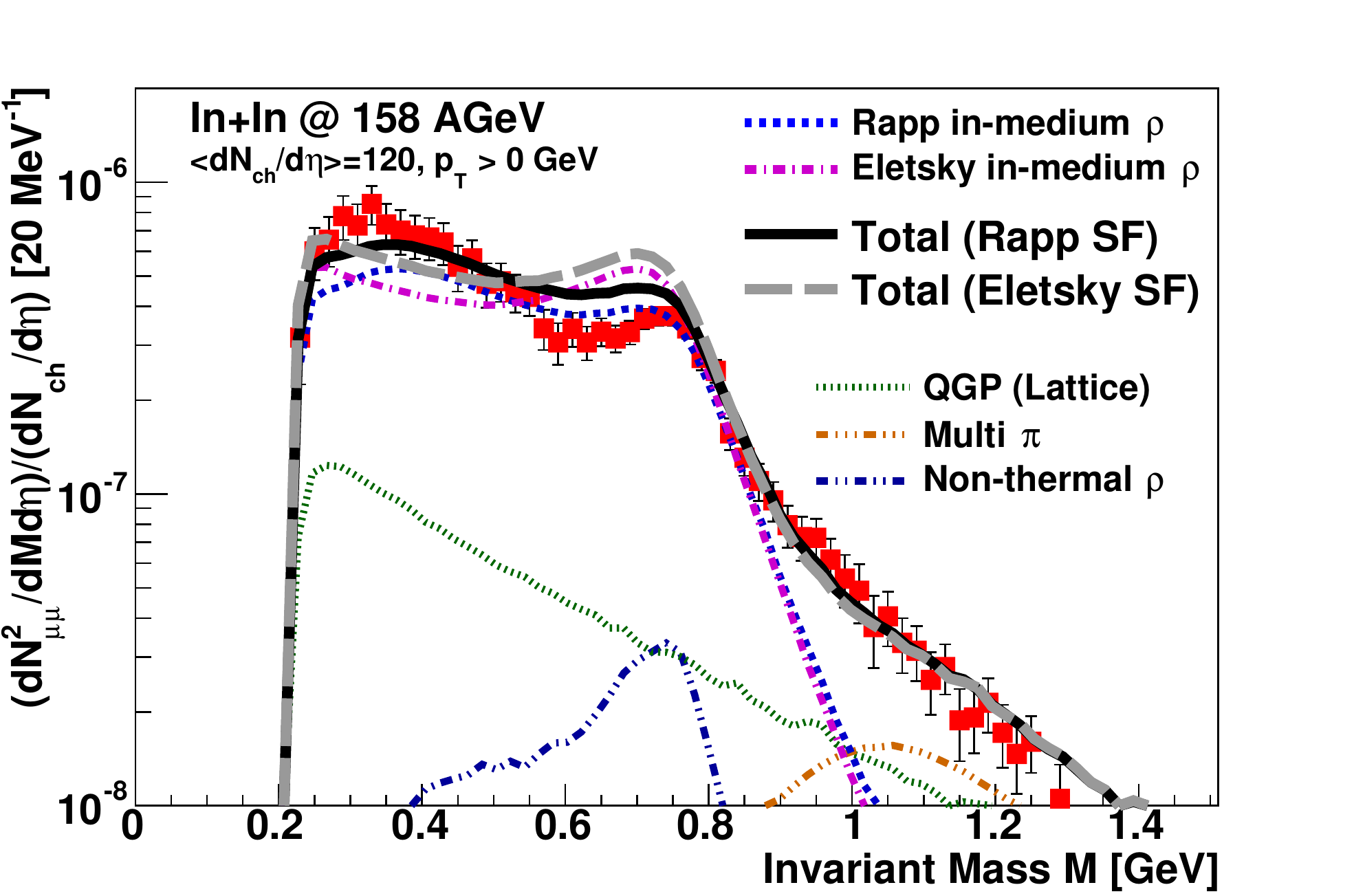}
\end{center} \vspace{-0.5cm} \caption{\label{fig3}Left: Di-muon spectrum
  for In+In collisions at 160~AGeV from hybrid-model simulations
  \cite{Santini:2011zw} in comparison to NA60 data
  \cite{Arnaldi:2008fw}.  Right: Di-muon spectrum for In+In collisions
  at 160~AGeV from coarse grained simulations in comparison to NA60 data
  \cite{Arnaldi:2008fw}.  Calculations with the Rapp-Wambach and Eletsky
  spectral functions are compared \cite{Endres:2015wea}.}
\end{figure}

\section{Dileptons} 

As a next step let us investigate the temperature and density evolution
with penetrating probes.  To this aim we compare coarse-grained
transport simulations to hybrid-model calculations to explore
differences due to the assumption of local equilibration and due to
different spectral functions.  Fig.  \ref{fig3} left shows the
hybrid-model calculation \cite{Santini:2011zw} (Eletsky spectral
function) for dimuon production in In+In reactions in comparison to the
NA60 data \cite{Arnaldi:2008fw}.  One observes that the hybrid model
provides a good description of the experimental data, however with a
slight overestimation of the yield around the $\rho$ peak.  We compare
this to the coarse-grained transport simulation \cite{Endres:2014zua} in
Fig.  \ref{fig3} (right).  In this case we show in addition a comparison
to the Rapp-Wambach spectral function \cite{rw99b,Rapp:1999ej,vanHees:2006ng}.
Generally we observe a very good description of the NA60 data
\cite{Arnaldi:2008fw}.  The main differences between both approaches
seem to be caused by the different spectral functions, with the
Rapp-Wambach spectral function providing a better description of the
data.  The hybrid-model results also show an excess above the data of
the $\rho$ around its pole mass. This mainly stems from the initial and
final stage which are handled by the hadronic cascade and where (in
contrast to the hydrodynamic phase) no explicit medium modifications of
the spectral shape are considered.
\begin{figure}[t]
\begin{center}
\includegraphics[width=0.45\textwidth]{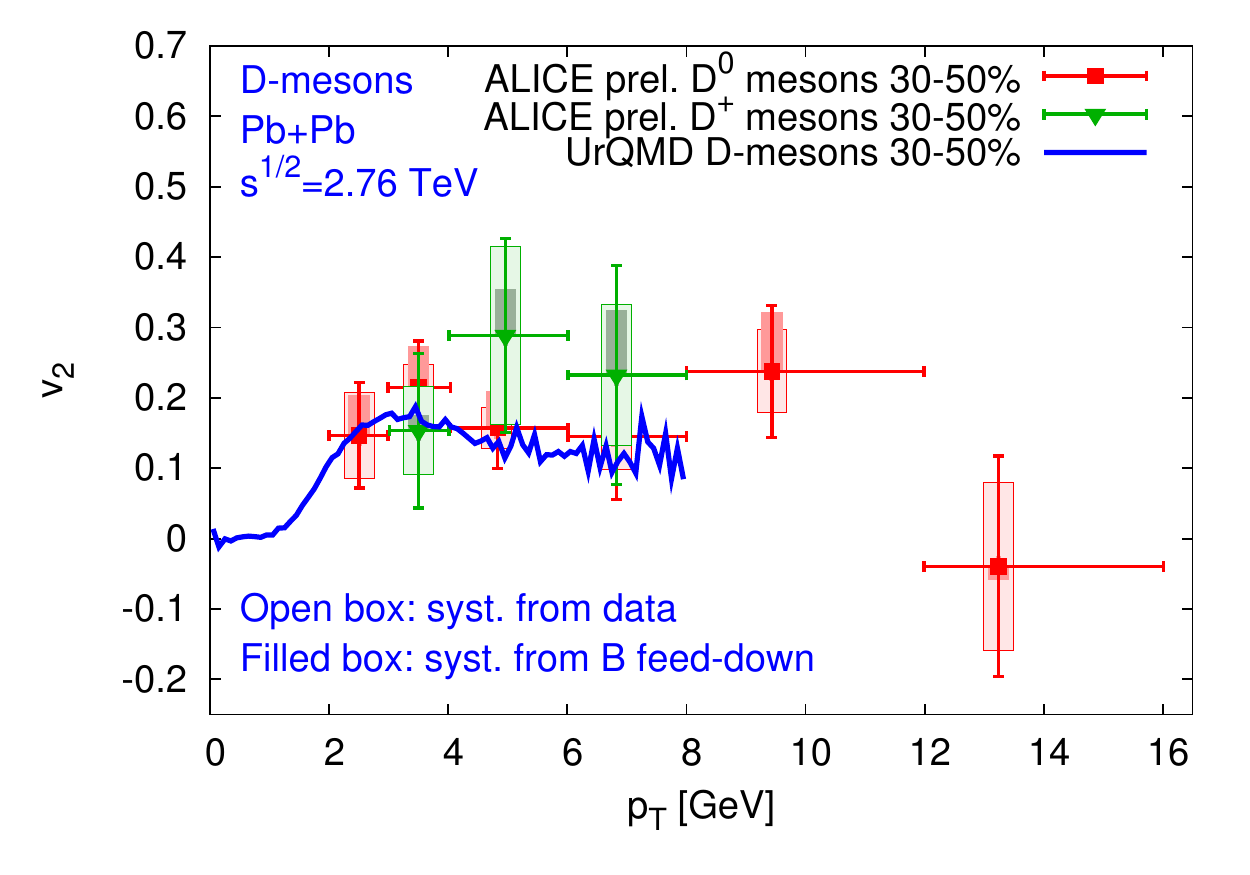}\includegraphics[width=0.45\textwidth]{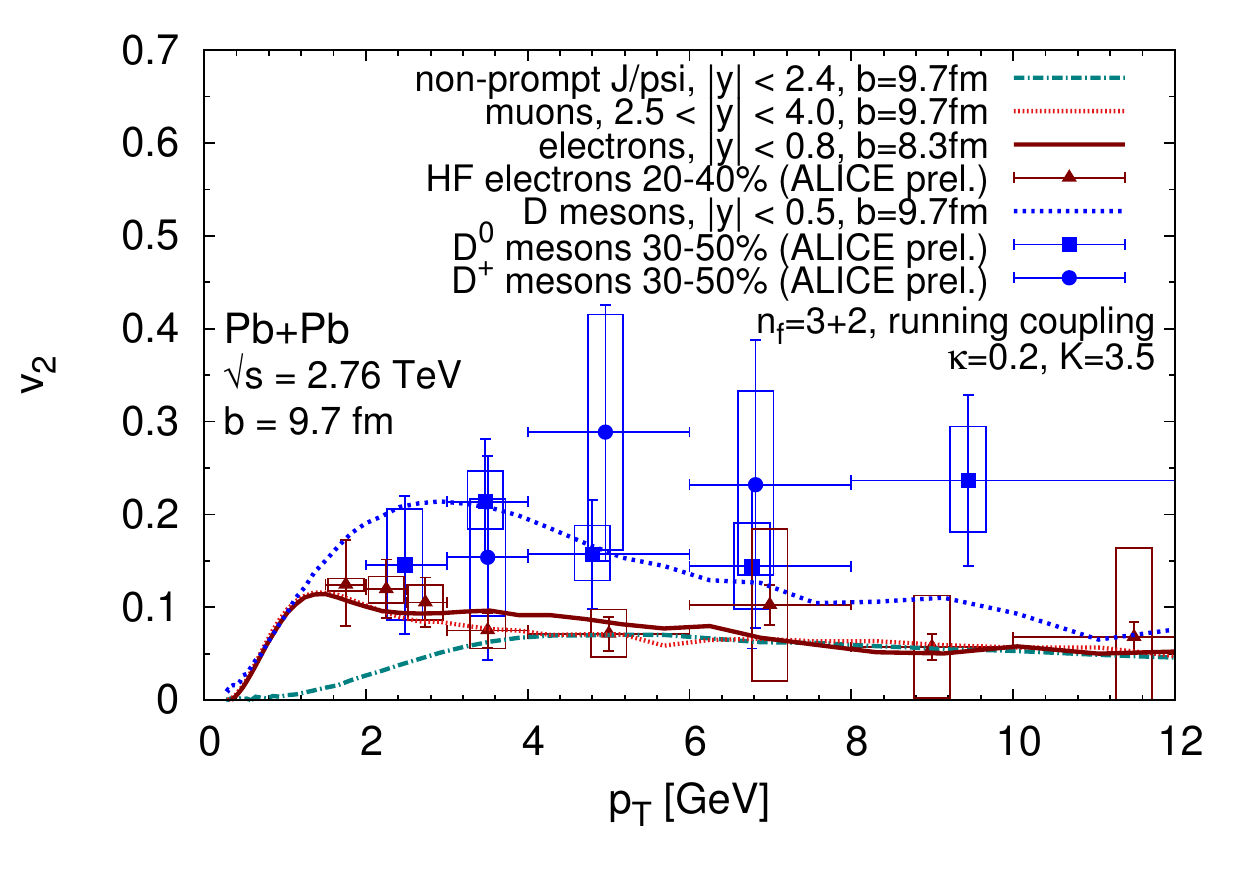}
\end{center} \vspace{-0.5cm} \caption{\label{fig4}Pb+Pb reactions at $\sqrt
{s_{NN}}=2.76$~TeV.  Models compared to ALICE data \cite{Ortona:2012gx}. 
Left: Hybrid model calculations with a Langevin treatment of the heavy quark
dynamics \cite{Lang:2012cx}.  Right: Parton cascade calculation.  Figure
taken from \cite{Uphoff:2012it}.}
\end{figure}

\section{Charm} An alternative way to explore the properties of the
created matter is to investigate its transport properties in terms of
drag and diffusion coefficients.  To this aim we model the propagation
of heavy quarks, i.e., charm quarks through the matter created in the
hybrid approach.  In Fig.  \ref{fig4} we compare the results for the
D-meson elliptic flow $v_2$ in Pb+Pb collisions at the LHC based on the
hybrid model \cite{Lang:2012cx} (left figure) with the results obtained
from a recent parton cascade \cite{Uphoff:2012it} study (right figure).
For both cases we observe a similar quality for the description of the
experimental data.  This indicates that the density and temperature
distribution and evolution in the hybrid approach and the parton cascade
approach seem to be very similar.  This may indicate that a substantial
amount of local equilibrium is achieved in such collision as the
dynamics of the heavy quarks does not seem to depend on the details of
the system evolution.

\section{Summary} 

In summary, hybrid models combining a hydrodynamic
simulation for the hot and dense stages of the reaction coupled to Boltzmann
dynamics for the early and late stage of the evolution provide an excellent
tool to investigate the properties of QCD matter created at SPS, RHIC and
LHC.  For these proceedings we have discussed the modifications of the
(anti-)proton yields due to absorption effects in the final state hadronic
afterburner.  We have then explored the temperature and density evolution by
means of dilepton radiation and charm dynamics.  In comparison to available
data and alternative approaches, we concluded that the hybrid approach
provides a reliable and sensible basis for these investigations.

\section{Acknowledgements}

M.\ B. wants to thank the organizers for an excellent meeting and many
fruitful and stimulating discussions. This work has been supported by
the Hessian LOEWE initiative through HIC for FAIR.


\begin{thebibliography}{99} 


\bibitem{Bass:1998ca}
  S.~A.~Bass {\it et al.},
  Prog.\ Part.\ Nucl.\ Phys.\  {\bf 41}, 255 (1998)
  [Prog.\ Part.\ Nucl.\ Phys.\  {\bf 41}, 225 (1998)]
  [arXiv:nucl-th/9803035].

\bibitem{Bleicher:1999xi}
  M.~Bleicher {\it et al.},
  J.\ Phys.\  {\bf G25}, 1859 (1999)
  [arXiv:hep-ph/9909407].

\bibitem{Petersen:2008dd}
  H.~Petersen, J.~Steinheimer, G.~Burau, M.~Bleicher and H.~Stocker,
  Phys.\ Rev.\  C {\bf 78}, 044901 (2008)
  [arXiv:0806.1695 [nucl-th]].

\bibitem{Bleicher:1998wd}
  M.~Bleicher {\it et al.},
  Nucl.\ Phys.\  A {\bf 638}, 391 (1998).

\bibitem{Huovinen:2012is}
  P.~Huovinen and H.~Petersen,
  arXiv:1206.3371 [nucl-th].

\bibitem{Becattini:2012sq}
  F.~Becattini, M.~Bleicher, T.~Kollegger, M.~Mitrovski, T.~Schuster and R.~Stock,
  Phys.\ Rev.\  C {\bf 85}, 044921 (2012)
  [arXiv:1201.6349 [nucl-th]].

\bibitem{Magas:2002ge}
  V.~K.~Magas, L.~P.~Csernai and D.~Strottman,
  Nucl.\ Phys.\  A {\bf 712}, 167 (2002)
  [arXiv:hep-ph/0202085].

\bibitem{Csernai:2005ht}
  L.~P.~Csernai, V.~K.~Magas, E.~Molnar, A.~Nyiri and K.~Tamosiunas,
  Eur.\ Phys.\ J.\  A {\bf 25}, 65 (2005)
  [arXiv:hep-ph/0505228].

\bibitem{Andrade:2005tx}
  R.~Andrade, F.~Grassi, Y.~Hama, T.~Kodama, O.~.~J.~Socolowski and B.~Tavares,
  Eur.\ Phys.\ J.\  A {\bf 29}, 23 (2006)
  [arXiv:nucl-th/0511021].

\bibitem{Hirano:2005wx}
  T.~Hirano and M.~Gyulassy,
  Nucl.\ Phys.\  A {\bf 769}, 71 (2006)
  [arXiv:nucl-th/0506049].

\bibitem{Nonaka:2005aj}
  C.~Nonaka and S.~A.~Bass,
  Nucl.\ Phys.\  A {\bf 774}, 873 (2006)
  [arXiv:nucl-th/0510038].

\bibitem{Werner:2010aa}
  K.~Werner, I.~Karpenko, T.~Pierog, M.~Bleicher and K.~Mikhailov,
  Phys.\ Rev.\  C {\bf 82}, 044904 (2010)
  [arXiv:1004.0805 [nucl-th]].

\bibitem{Steinheimer:2012rd}
  J.~Steinheimer, J.~Aichelin and M.~Bleicher,
  Phys.\ Rev.\ Lett.\  {\bf 110}, 042501 (2013)
  [arXiv:1203.5302 [nucl-th]].

\bibitem{Pan:2014caa}
  Y.~Pan and S.~Pratt,
  Phys.\ Rev.\ C {\bf 89} (2014) 4,  044911.

\bibitem{Becattini:2012xb}
  F.~Becattini, M.~Bleicher, T.~Kollegger, T.~Schuster, J.~Steinheimer and R.~Stock,
  arXiv:1212.2431 [nucl-th].

\bibitem{Becattini:2014hla}
  F.~Becattini, E.~Grossi, M.~Bleicher, J.~Steinheimer and R.~Stock,
  Phys.\ Rev.\ C {\bf 90} (2014) 5,  054907
  [arXiv:1405.0710 [nucl-th]].

\bibitem{Li:2008qm}
  Q.~f.~Li, J.~Steinheimer, H.~Petersen, M.~Bleicher and H.~Stocker,
  Phys.\ Lett.\  B {\bf 674}, 111 (2009)
  [arXiv:0812.0375 [nucl-th]].

\bibitem{Santini:2011zw}
  E.~Santini, J.~Steinheimer, M.~Bleicher and S.~Schramm,
  Phys.\ Rev.\  C {\bf 84}, 014901 (2011)
  [arXiv:1102.4574 [nucl-th]].

\bibitem{Arnaldi:2008fw}
  R.~Arnaldi {\it et al.}  [NA60 Collaboration],
  Eur.\ Phys.\ J.\  C {\bf 61}, 711 (2009)
  [arXiv:0812.3053 [nucl-ex]].

\bibitem{Endres:2014zua}
  S.~Endres, H.~van Hees, J.~Weil and M.~Bleicher,
  arXiv:1412.1965 [nucl-th].

\bibitem{rw99b}
R.~Rapp, J.~Wambach, Eur. Phys. J. A \textbf{6}, 415 (1999).

\bibitem{Rapp:1999ej}
  R.~Rapp and J.~Wambach,
  Adv.\ Nucl.\ Phys.\  {\bf 25}, 1 (2000)
  [arXiv:hep-ph/9909229].

\bibitem{vanHees:2006ng}
  H.~van Hees and R.~Rapp,
  Phys.\ Rev.\ Lett.\  {\bf 97} (2006) 102301
  [hep-ph/0603084].

\bibitem{Endres:2015wea}
  S.~Endres, H.~van Hees, J.~Weil and M.~Bleicher,
  arXiv:1502.01948 [nucl-th].

\bibitem{Lang:2012cx}
  T.~Lang, H.~van Hees, J.~Steinheimer and M.~Bleicher,
  arXiv:1211.6912 [hep-ph].

\bibitem{Uphoff:2012it}
  J.~Uphoff, O.~Fochler, Z.~Xu and C.~Greiner,
  Nucl.\ Phys.\  A {\bf 910-911}, 401 (2013)
  [arXiv:1208.1970 [hep-ph]].

\bibitem{Ortona:2012gx}
  G.~Ortona  [ALICE Collaboration],
  arXiv:1207.7239 [nucl-ex].


\end{thebibliography}
\end{document}